# Conference summary: Workshop on Precision Astronomy with Fully Depleted CCDs (2014)

**J. A. Tyson**[a,*]

[a] *University of California,*
  *Davis, CA, USA*
  *E-mail*: tyson@physics.ucdavis.edu

ABSTRACT: Thick fully depleted CCDs, while enabling wide spectral response, also present challenges in understanding the systematic errors due to 3D charge transport. This 2014 Workshop on Precision Astronomy with Fully Depleted CCDs covered progress that has been made in the testing and modeling of these devices made since a workshop by the same name in 2013[1]. Presentations covered the science drivers, CCD characterization, laboratory measurements of systematics, calibration, and different approaches to modeling the response and charge transport. The key issue is the impact of these CCD sensor features on Dark Energy science, including astrometry and photometry. Successful modeling of the spatial systematics can enable first order correction in the data processing pipeline.

---

[1] http://iopscience.iop.org/1748-0221/focus/extra.proc34

**Contents**

# 1. Introduction

In astronomical optical survey applications, mosaics of CCDs are currently the detector of choice. This was not always the case. Up to the end of the 1970s photographic plates coupled with a wide field corrector offered several degree scale imaging per exposure. The Palomar sky survey is an example. But there were four problems with photographic plates: (1) low quantum efficiency (typically 5% or less), (2) one had an entirely different detector for each exposure, (3) the resulting stored image was analog (the plates had to be digitized with a scanning spot), and (4) the dynamic range of the photographic process was small (about 8 bits). The tiny 3x4mm CCDs we played with in the mid 1970's were thus a curiosity, and it was not obvious that they would supplant photographic plates for wide-field imaging anytime soon. Even though the first CCDs were noisy, two properties were quite interesting, particularly for imaging of small fields of view: (1) the QE was ten times that of the best hypered plates, and (2) it was the same detector on every exposure. Of these two attractive properties, the second ultimately became crucial for imaging of faint surface brightness. The CCD shift-and-stare imaging we developed in the early 1980s made use of this fact, enabling the separation of systematic spatial features on the CCD from the true astronomical scene.

 As CCD technology improved it became clear that they would replace photographic plates. Buried channel operation reduced many of the traps, and lower noise on-chip amplifiers drove the noise down from 1000$e$ per read to a few $e$ per pixel per read. The size of CCDs grew, and CCD mosaic cameras covering the same real estate as the old photographic plates were built. Back-illuminated CCDs increased the QE and UV response. These devices were partially depleted and relatively thin (20-40 microns). To first order, charge transport was not an issue; the photoelectrons followed the back bias field; photons arriving at position *xy* were reported on readout at the same position *xy*. In recent years thick fully depleted CCDs have been designed which offer wider spectral response from 300 nm to beyond 1 micron. This effort was led by Steve Holland at LBL; those 250 micron thick devices are used on the Dark Energy Camera. Indeed he told us at the workshop about a new CCD with one electron read noise! The LSST project designed its own thick fully depleted CCDs; the thickness is limited to 100 microns due to the fast f/1.2 optical beam of the camera. In both cases, the pixels are small compared with the thickness, making a pixel that resembles a skyscraper! Photoelectrons on their way from the photoconversion site to the front-side gates are subject to transverse electric fields, giving rise to astrometric and photometric systematics. The measurement and proposed cure of these effects was the subject of this most recent workshop.

## 2. Science

### 2.1 Science drivers of detector performance

The extended red response of thick fully depleted CCDs enables photometric redshift measurements on high redshift galaxies. The thick design with small pixels (typically 10 microns) presents challenges for the precision astrometric performance specifications. There are two science drivers of good astrometric precision: studies of stellar motion (5 milli arcsecond stellar astrometry), and weak gravitational lensing. The weak lens shear error is the derivative of the astrometric error.

Rachel Mandelbaum told us in her review talk about the weak lensing shear reconstruction challenges. There are both multiplicative and additive shear systematics, and they can both be affected by CCD astrometric errors. While the weak lens shear around a cluster of galaxies is a few percent, the cosmic shear caused by fluctuations in cosmic dark matter density is one to two orders of magnitude smaller. Cosmic shear is a powerful probe of cosmology, and the physics of dark energy. The shapes of billions of faint galaxies will be measured, from which one can derive the shear field – only if instrumental and atmospheric effects can be corrected. For the next generation cosmic shear experiment PSF shear systematics must be kept below 3E-4. These astrometric errors can occur either from atmospheric smearing of the image during exposure, or CCD-based errors. The shapes of faint galaxies must be corrected for these effects. What is needed is a reliable estimate of the PSF at every point in the image. Finally, the size of pixels is also changed by these distortions, causing photometric systematics.

### 2.2 Stars as PSF calibrators

Luckily there are about 100 stars on each CCD that are bright enough to be used as local estimates of the PSF. However, the peak flux in these stellar images is typically higher than that of the faint galaxies whose shapes we need to measure. The "brighter-fatter" effect discussed below causes a systematic shape error of these stars relative to the faint PSF appropriate to the faint galaxies. Moreover, the pixels in the CCD present a distorted (sheared) image of the sky because of charge transport effects in the thick CCD. These effects and their modelling were a focus of the workshop. Were we to ultimately correct for these astrometric effects, there remains another challenge: extrapolating the corrected stellar PSF to the position of the galaxy. This requires a more detailed model of the spatial dependence of the PSF over the CCD. Forward modelling, using all the information available, is a good approach as long as model bias can be quantified by simulations. An example is *sFIT*, an iterative Bayesian algorithm by James Jee which won the recent blind shear estimation challenge GREAT3 [1]. For cosmology with shear, the shear estimation is a statistical problem with error estimates from large covariance calculations: Bob Armstrong reminded us of Michael Schneider's hierarchical probabilistic model of shear inference which can offer reduction in shear bias while giving improved knowledge of latent classes of galaxy ellipticity variance. Josh Meyers told us about an atmospheric effect due to observing away from the zenith, PSF chromaticity, and the likely prospect of correcting for it at the catalog level – given enough observational data on the astrometric residuals of stars of different color observed at a range of airmass.

**2.3 Thick CCD cameras in use**

We heard results from three teams who have recent experience with CCD camera systematics analysis from observational data. First, Satoshi Miyazaki told us about his impressive Hyper Suprime-Cam and showed us a zoomed image into a distant cluster of galaxies taken from the HSC-Subaru sky survey which is under way. This camera uses thick CCDs and the data reduction challenges cover the gamut from optics ghosts to CCD effects. There will be lessons learned from analysis of repeated deep imaging with HSC. While the Pan-STARRS camera uses thinner CCDs, Gene Magnier told us about their complicated systematics analysis issues. Finally, data from the 250 micron thick CCDs in Dark Energy Camera on the Blanco telescope was shown in several talks (and discussed below). The charge transport effects are revealed at high S/N in the DEC because of the thickness of their CCDs. There already are lessons learned from those data.

## 3. The problem

The challenges of 3D charge transport in fully depleted CCDs was reviewed by Chris Stubbs at the previous workshop [2]. As Pierre Astier and Andrei Nomerotski reminded us in their introductory talks, electron optics becomes an issue in thick fully depleted CCDs. Even though there is a high back bias field, there are transverse electric fields throughout the CCD which cause charge transport complications. Since the drift path length of an electron in fully depleted regions is much larger than a pixel, these transverse fields cause a mapping between the input focal plane and the CCD output gate structure. There are three separate spatial regions where photoelectrons may be deflected from their intended path: (1) near the edges of the CCD where there are fringing fields, (2) near the front-side electrodes and channel stops, and (3) throughout the fully depleted thickness of the device. So-called "tree rings" caused by silicon boule growth impurity variations are an example of the latter. Generally, these charge transport effects are functions of applied voltages, CCD design, impurity variations (including possible non fully depleted regions), optical flux induced space charge effects, temperature, and wavelength.

Wavelength dependence of charge transport arises mainly from the dependence of the mean free path of a photon as the wavelength corresponding to the band gap is approached. The hope of course is that these effects are constant with voltage and temperature for a given CCD, so that they may be characterized, modelled, calibrated, and corrected in the data. We heard a lot about tree rings. Yuki Okura showed attempts to model them, and Chris Walter showed us his simulations using the LSST end-to-end simulator *phosim*.

There are of course other "features" in CCDs – unrelated to these charge transport issues -- arising from their architecture and exposure to radiation. Paul O'Connor told us about cross-talk in the highly segmented LSST CCDs, and our solution for disambiguating them using the technique of dithering by rotation. Space applications of solid state detectors is always a challenge for long missions in radiation environments; Douglas Jordan told us about the radiation induced CTE in the Euclid mission devices.

## 4. Calibration

No matter how well we characterize our detectors in the lab, there is a need for calibration during operations. However it is asking for trouble to depend only on in-dome calibration or on-sky calibration, given the dimensionality of our thick CCD effects. In-dome calibration will also be necessary if for no other reason that there can be drift with time in external QE and optical filter and telescope optics throughput. Robert Lupton reviewed for us the clever calibration scheme combining Chris Stubbs' special projection hardware with an iterative solution to the ill posed problem of calibrating photometry when you don't know all the terms in the flux integral apriori. I think this may work at the catalog level as intended, separating out ghost and scattered light, given enough projector data. The projection system for LSST will scan daily vs wavelength and can project many spots onto the focal plane. It should calibrate the system throughput below the atmosphere. A small spectroscopic patrol telescope (together with extra-dome measurements of water vapour and pressure vs height) can complete the calibration through the atmosphere.

We heard from Bob Armstrong about the DES on-sky PSF calibration program, which appears to be working well. All data from stars in all images is analysed for both astrometric and photometric residuals. We were again reminded of ghosts – in particular chromatic ghosts due to optical filter edge response -- in a talk by Nicolas Regnault on their analysis of the CFHTLS data. Finally, William Wester told us about the DES experience of the effects of chromatic ghosting, and the spectrophotometric calibration system for DECam.

## 5. Lab measurements

Much of what we know about the LSST CCDs come from lab tests at Harvard, BNL, UC Davis, and Paris which have been in place for some time. Most of these test setups check for pixel gain, response, and correlations using flat field illumination and projected single spots, as well as flatness and read noise measurements. These systems are intended to be used in device acceptance testing. There are iron-55 tests for charge diffusion; and Merlin Fisher-Levine even told us how cosmic ray muons can be used! The most ambitious test system illuminates an entire CCD with a facsimile of the LSST camera f/1.2 beam. Such a system can be used to fully characterize the low level charge transport effects which could impact weak lens science. While outside of the initial device acceptance testing, these studies are necessary if the science goals of LSST are to be achieved. In short, we are not done when the detectors are delivered.

Andrew Bradshaw discussed measurements using this system at UC Davis, in which a projected grid of 40,000 "stars" was dithered many times in the plane of the CCD with sub-pixel sampling, as a function of intensity. Pixel astrometric errors were seen at CCD serial register edges and at the bloom stop in the middle. Of course the brighter-fatter effect (discussed below) is also seen, but the news was the discovery of intensity dependent astrometric errors near the bloom stop. Sub-pixel investigations of channel stop effects are planned using 3 micron projected hole arrays. Mask errors are known to exist on and CCD, and enough data of this sort has the potential of solving for the mask errors and the charge transport map separately.

## 6. Modeling charge transport

### 6.1 Brighter-fatter effect

This effect is due to space charge near the gate structure, and was the first obvious charge transport issue discovered. As mentioned above, its largest impact is mis-estimation of weak lens shear (multiplicative systematic). It must be corrected to a part per thousand. The brighter-fatter (BF) effect appears linear in photon flux (and photoelectrons per pixel) as expected. Augustin Guyonnet told us about an empirical method for estimating the effect. Building on the clever idea of Pierre Antilogus, pixel correlations in differences of flat fields are used to extract the BF amplitudes vs intensity. The BF effect is anisotropic, due to the channel stops; 50 simultaneous equations are solved, and a correction accuracy of a few percent has been achieved. While this empirical method using all pixels in a correlation study is capable of giving mean BF vs intensity fits on a per CCD basis, it cannot produce detailed maps of the BF effect which will likely be necessary if we want to correct for spatial variations.

The BF effect is a function of voltages (back bias and clocks) as expected. The space charge model would predict that the BF amplitude should depend only on the collected charge per pixel. Daniel Gruen told us that it is independent of photon rate, thus ruling out more complicated physics. However, he also told us that it appears to be CCD lot dependent. This raises two issues: first we must characterize all production lots, and second we probably have to worry about variations in doping of the channel stops from lot to lot and perhaps even within a device.

### 6.2 More general spatial systematics

At the last PACCD meeting in 2013 (and in a poster at this meeting) Andy Rasmussen presented a more general analysis of spatially dependent charge transport in thick CCDs in which he models electric fields using distributed dipole charges. This model for the LSST CCDs has been incorporated in the *phosim* image simulator. Chris Walter alerted us to the difference between the Guyonnet approach and this more general approach which is based on details of the polysilicon structures and voltages. An intermediate approach is to take the latter as variables to be fit by the data, at a loss of precision due to the higher dimensionality of the fit. In fact there is another related approach in which Craig Lage has been modelling the 3D spatially dependent fields at sub-pixel resolution (and the resulting charge transport) by solving Poisson's equation for the known voltages and CCD geometry. This geometrical information must include not only the layout of the polys but also the geometry of non-depleted regions on sub-pixel scales. One thing is clear: if we want the best model we need to know some basic information from the CCD vendors, sufficient to give these boundary conditions for the Poisson solver.

## 7. Correction of systematics

Assuming we can adequately model the various astrometric and photometric effects, including mask errors, then we can in principle correct the CCD data in the data reduction pipeline. The four corners of each pixel are mapped in the simplest scheme. The shape of pixels is also affected, and a more precise mapping would include the pincushion like pixel distortions caused by the field lines near the channel stops. The mapping model would have to scale to all relevant voltages, wavelengths, and temperatures. Finally, this pixel mapping will also be a function of sky background. All this is known either from camera data or from the sky image itself.

How well do we have to do? A toy dithered imaging simulation of 100 visits to a field using the known edge and bloom-stop astrometric errors on our prototype LSST CCDs suggests that a factor of ten better correction than that currently being made by DES may be sufficient. The details of the dithering matter: dithering by camera rotation in addition to xy dithering is necessary.

There was an atmosphere of cautious optimism regarding the likelihood of ultimately correcting for the static and dynamic charge transport effects in thick fully depleted CCDs. The way forward must include lab measurements on a variety of spatial scales and over a range of clocking and back bias voltages. A physically motivated model fit to the lab data would be preferable to a heuristic model.